\documentclass[final,5p,times,twocolumn]{elsarticle}
\pdfoutput=1

\usepackage{graphics}
\usepackage{lineno}
\usepackage{amsmath}
\usepackage{amssymb}
\newcommand{\DBD}{0$\nu$DBD}
\newcommand{\PO}{$^{210}$Po}
\newcommand{\TEO}{$\mathrm{TeO}_2$}
\newcommand{\ZNMO}{$\mathrm{ZnMoO}_4$}
\newcommand{\TL}{$^{208}\mathrm{Tl}$}
\newcommand{\THO}{$^{232}\mathrm{Th}$}
\newcommand{\Cuore}{CUORE}
\newcommand{\MC}{Monte Carlo}
\newcommand{\ckky}{\un{c/(keV\;kg\;y)}}

\providecommand*{\un}[1]{\ensuremath{\mathrm{~#1}}}

\journal{Astroparticle Physics}
\begin{document}

\begin{frontmatter}

\title{ZnMoO$_4$: a promising bolometer  for neutrinoless double beta decay searches}

\author[LBL]{J.W.~Beeman}
\author[INFNRM1,UNIROMA1]{F.~Bellini}
\author[INFNMIB,UNIMIB]{S.~Capelli}
\author[INFNRM1,UNIROMA1]{L.~Cardani}
\author[UNIAQ]{N.~Casali}
\author[INFNRM1]{I.~Dafinei}
\author[INFNGE,UNIGE]{S.~Di~Domizio}
\author[INFNRM1,UNIROMA1]{F.~Ferroni}
\author[NOVO]{E.~N.~Galashov}
\author[INFNMIB,UNIMIB]{L.~Gironi}
\author[INFNRM1]{F.~Orio}
\author[INFNMIB]{L.~Pattavina}
\author[INFNMIB]{G. Pessina}
\author[INFNRM1,UNIROMA1]{G.~Piperno}
\author[INFNMIB]{S. Pirro \corref{cor1}}\ead{Stefano.Pirro@mib.infn.it}
\author[NOVO]{V.~N.~Shlegel}
\author[NOVO]{Ya.~V.~Vasilyev}
\author[INFNRM1]{C.~Tomei}
\author[INFNRM1]{M.~Vignati}

\cortext[cor1]{Corresponding author}

\address[LBL]{Lawrence Berkeley National Laboratory , Berkeley, California 94720, USA}
\address[INFNRM1]{INFN - Sezione di Roma 1 I 00185 Roma - Italy}
\address[UNIROMA1]{Dipartimento di Fisica - Universit\`{a} di Roma La Sapienza I 00185 Roma - Italy}
\address[INFNMIB]{INFN - Sezione di Milano Bicocca I 20126 Milano - Italy}
\address[UNIMIB]{Dipartimento di Fisica - Universit\`{a} di Milano Bicocca I 20126 Milano - Italy}
\address[UNIAQ]{Dipartimento di Fisica - Universit\`{a} degli studi dell'Aquila  I 67100  L'Aquila - Italy}
\address[INFNGE]{INFN - Sezione di Genova   I 16146 Genova - Italy}
\address[UNIGE]{Dipartimento di Fisica - Universit\`{a} di Genova  I 16126 Genova - Italy}
\address[NOVO]{Nikolaev Institute of Inorganic Chemistry - SB RAS, 630090 Novosibirsk - Russia}

\date{\today}

\begin{abstract}
We investigate the performances of two \ZNMO\ scintillating crystals operated as bolometers,
in view of a next generation experiment to search the neutrinoless double beta decay of $^{100}$Mo.
We present the results of the  $\alpha$ vs $\beta/\gamma$  discrimination, obtained through the 
scintillation light as well as through the study of the shape of the thermal signal alone.
The  discrimination capability obtained at the 2615 keV line of \TL\ is   8 ~$\sigma$,  using the heat-light scatter plot, while 
it exceeds   20 ~$\sigma$ using  the shape  of the thermal pulse alone. 
The achieved FWHM energy resolution ranges from 2.4 keV (at 238 keV) to 5.7 keV (at 2615 keV).
The internal radioactive contaminations of the  \ZNMO\ crystals were evaluated through a 407 hours background measurement. 
The obtained limit is $< 32~\mu$Bq/kg for $^{228}$Th and $^{226}$Ra. These values were used for a \MC\ simulation aimed at evaluating
the achievable background level of a possible, future array of enriched Zn$^{100}$MoO$_4$ crystals. 
\end{abstract}


\end{frontmatter}

\section{Introduction}

Double Beta Decay (DBD) searches became of critical importance after the discovery of  neutrino oscillations. Plenty of experiments 
are now in the construction phase and many others are in R\&D  phase~\cite{Avignone:2007fu,Bara-2011,EXO-2011,KamZen-2012}.
The main challenges for all the different experimental techniques are the 
same~\cite{Pirro:2006tv}: i) increase the active mass, ii) decrease the background, and iii)  increase the energy resolution.

Thermal bolometers are  ideal detectors for this survey: crystals  can be grown with a variety of  interesting DBD-emitters and multi 
kg detectors can be operated with excellent energy resolution~\cite{massive-tellurium}: perhaps most critical to the next generation experiments. 

The Cuoricino experiment~\cite{Andreotti:2010vj}, an array of 62 \TEO\ crystal bolometers, demonstrated not only the power of 
this technique, but also that the main source of background for these detectors arises from surface contaminations of radioactive 
$\alpha$-emitters. $\alpha$ particles, emitted from radioactive contaminations located on the surfaces of the detector or
of passive elements facing them, can lose part of their energy in a few $\mu$m and deposit the rest in the crystal bolometer.
This produces an essentially flat background starting from the Q-value of the decays (several MeV) down to low energies, completely 
covering, therefore, the region of  the Q$_{\beta\beta}$ values.  
Moreover  simulations show that this contribution will largely dominate the expected background of  the CUORE 
experiment~\cite{ACryo,Pavan:2008zz} in the region of interest, since there is no possibility to separate this background from the two DBD electrons. 
The natural way to discriminate this background is to use a scintillating bolometer~\cite{Pirro:2005ar}. 
In such a  device the simultaneous and independent readout of the heat and of the scintillation light permits to discriminate events 
due to $\beta/\gamma$, $\alpha$ and neutrons  thanks to  their different scintillation yield.
Moreover,  if the crystal is based on a DBD emitter whose transition energy exceeds the 2615 keV $\gamma$-line of \TL~,~the 
environmental background due to natural $\gamma$'s will decrease abruptly.

$^{100}$Mo is a very interesting $\beta\beta$-isotope because of its large transition energy $Q_{\beta\beta}=3034.4 $~keV~\cite{Rahaman:2007ng} 
and a considerable natural isotopic abundance $\delta=9.67\%$~\cite{Boh05}.
Several inorganic scintillators containing molybdenum were developed in the last years. \ZNMO\ ~\cite{Ivle08} is an example of crystal 
recently  tested as a cryogenic detector giving very promising results~\cite{Gironi:2010hs}.

In this work we present the results obtained with two \ZNMO\ crystals of superior quality with respect to the sample previously studied.
This work is focused on the  recent observation~\cite{Arnaboldi:2010gj}  that the  thermal signal induced by $\alpha$ and $\gamma$/$\beta$
particles shows a slightly different time development. 
This feature  seems to  be explained~\cite{Gironi:2011js} by the relatively long scintillation decay time  (of the order of 
hundreds of $\mu$s) observed in  some scintillating crystals (e.g. molybdates). This long decay, combined with an high percentage 
of non-radiative  de-excitation of  the scintillation channel, will transfer phonons (i.e. heat) to the crystal.
This extremely tiny, but measurable, time dependent phonon release has a different absolute value for  isoenergetic $\alpha$ and 
$\beta/\gamma$ particles due to their different scintillation yield.

The possibility to have a bolometer in which the $\alpha$ background is identified \emph{without} the need of an additional  light
detector is particularly appealing since it translates in a simplified set-up.

After an exhaustive description of the $\alpha$ vs $\beta/\gamma$  discrimination power, presented in Sec.~\ref{sec:discrimination}, in
Sec.~\ref{sec:background-simulation} we present  a detailed \MC\ simulation of the background performances of a possible 
future experiment  based on a \ZNMO\ crystal array.
Within the Lucifer Project~\cite{Ferroni2010}, an array of enriched Zn$^{82}$Se crystals operated as scintillating 
bolometers, there is  also the option   of having part of the detectors made of   $^{100}$Mo enriched crystals.
The present work, though based  on  small size crystal samples,  shows  the performance that could be reached by a small scale 
DBD decay experiment based on Zn$^{100}$MoO$_4$ crystals.

\section{Experimental details}
\label{sec:Experimental-details}
High quality \ZNMO\ crystals   were developed in the Nikolaev Institute of Inorganic Chemistry (NIIC, Novosibirsk, Russia).
Starting material for the crystal growth were High Purity ZnO (produced by Umicore)  and MoO$_3$,  synthesized  by NIIC.  
Crystals up to 25 mm in diameter and 60 mm in length were grown by the low-thermal-gradient Czochralski technique in a 
platinum crucible with a size of $\oslash~40$ ~mm $\times 100$ mm ~\cite{Galashov-2010}.

In this paper we present the result of two separate runs, both carried out in an Oxford 200 $^3$He/$^4$He dilution refrigerator 
located deep underground in the Laboratori Nazionali del Gran Sasso (average depth$\approx$3650 m w.e.~\cite{Ambrosio-98}).

The first run was dedicated to the study of the discrimination capability of \ZNMO\ bolometers,
while the second one was devoted to the evaluation of the internal contaminations of the sample.

In the first run  two different \ZNMO\ crystals were tested in order to study the reproducibility of the background rejection over 
different samples.
The samples were a 27.5 g colourless  cylinder ($\oslash~18.5$~mm $\times 22.3$ mm) and a 29.9 g slightly orange 
parallelepiped (28.5$\times$18.4$\times$13.2 mm$^3$).

\begin{figure}[t]
\centering
\includegraphics[width=0.48\textwidth,clip=true]{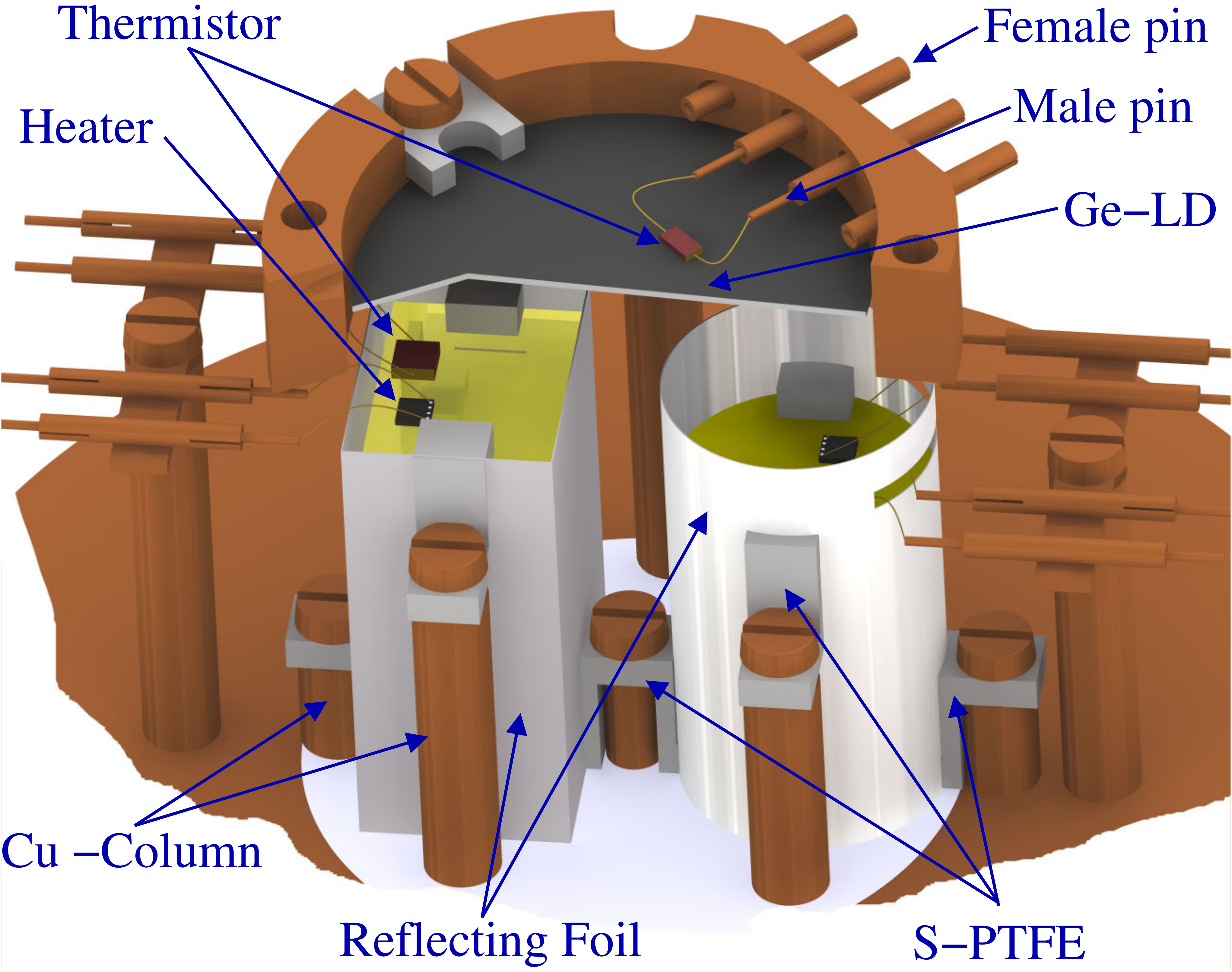}
\caption{Set-up of the detectors. The ball-bonded Au  wires are crimped into  ``male'' Cu tubes (pins) and inserted into ground-insulated ``female'' Cu tubes. 
Custom wires from detectors towards cryostat are not drawn. A section of the light detector is not drawn for a better understanding.}
\label{fig:setup}      
\end{figure}

The \ZNMO\ crystals were held by means of four S-shaped PTFE supports fixed to cylindrical Cu columns. 
They  were surrounded (with no direct contact) by a plastic  reflecting sheet (3M  VM2002).
The Light Detector (LD)~\cite{Pirro-LD-2006}  is constituted by a  36 mm diameter, 1 mm thick pure Ge crystal absorber. 
The Ge wafer is heated up by the  absorbed photons  and the temperature variation is proportional to the scintillation signal.  
The set-up of the detectors is schematized in Fig.~\ref{fig:setup}.  

The temperature sensor of the \ZNMO\ crystals is a 3x3x1 mm$^3$ Neutron Transmutation Doped (NTD) germanium thermistor, the same used 
in the Cuoricino experiment. It is thermally coupled to the crystal via 9 glue spots of $\approx$~0.6 mm diameter and $\approx$50 $\mu$m height.
The temperature sensor of the LD has a smaller volume (3x1.5x0.4~mm$^3$) in order to decrease its heat capacity, increasing therefore its 
thermal signal.
A resistor of  $\approx$300 k$\Omega$, realized with a heavily doped  meander on a 3.5 mm$^3$ silicon chip, is attached to  
each absorber and acts as a heater to  stabilize the gain of the bolometer~\cite{stabilization,Arnaboldi:2003yp}.
The details of the electronics and the cryogenic facility can be found elsewhere \cite{NIMA-2006-B,NIMA-2006-C,NIMA-2004}.

The heat and light pulses, produced by a particle interacting in  the absorber and transduced in a voltage pulse by 
the NTD thermistors, are amplified and fed into a 18 bit NI-6284 PXI ADC unit.
The trigger is software generated on each bolometer and when it fires waveforms 0.6~s long, sampled at 2 kHz,  are  saved on disk. 
Moreover, when the trigger of a \ZNMO\ crystal fires, the corresponding waveform from the LD is recorded,  irrespective of its trigger.

As the main goal of the measurements was to  test  the discrimination capability of the detectors
between $\alpha$ and $\beta/\gamma$ events, a $^{238}$U/$^{234}$U $\alpha$ source was faced to the crystals, on the opposite side 
with respect to the LD.
The  source  was covered with a 6 $\mu$m thick Mylar film, in order to ``smear'' the  $\alpha$'s energies down to the Q$_{\beta\beta}$ energy region.
The $\gamma$ calibration of the \ZNMO\ crystals is performed through a movable  a \THO\ source inserted between the Dewar housing the cryostat 
and the external lead shield.
The energy calibration of the LD is achieved thanks to a $^{55}$Fe X-ray source, producing  two X-rays at 5.9 and 6.5~keV,  faced  to the 
LD.

In the second run only the best performing bolometer (the cylindrical crystal) was characterized. The $\alpha$ source was removed and 
the crystal was surrounded by several layers of ultrapure polyethylene in order to shield the bolometer against the surface contamination of the 
copper structure.
In this measurement the lack of space prevented us from mounting the LD. 
However this was not a problem, as the previous run convincingly demonstrated that 
the pulse shape analysis can provide an extremely good $\alpha$ background rejection without the need for the light detection.

\subsection{Data Analysis}
\label{subsec:da}
To maximize the signal to noise ratio, the pulse amplitude  is estimated by means of an optimum filter technique~\cite{Gatti:1986cw,Radeka:1966}.
The filter transfer function is built from  the ideal  signal shape $s(t)$ and  the noise power spectrum $N(\omega)$.
$s(t)$ is estimated by averaging a large number of triggered pulses (so that stochastic noise superimposed  to each pulse averages to zero) 
while $N(\omega)$  is computed  averaging the power spectra of randomly acquired waveforms not containing pulses.
The amplitude of a signal is estimated as the maximum of the filtered pulse. This procedure is applied 
for the signal on the \ZNMO\ bolometer.
The amplitude of the light signal, instead, is estimated from the value of the filtered waveform at a fixed time delay with respect to 
the signal of the \ZNMO\ bolometer, as described in detail in Ref.~\cite{Piperno:2011fp}.
The detector performances  are reported in Table \ref{Table:parameters_measurement}. 
The baseline resolution, FWHM$_{base}$, is governed by the noise fluctuation at the filter output, and does not depend on the absolute pulse 
amplitude. 
The rise ($\tau_{R}$) and decay times ($\tau_{D}$) of the pulses  are computed as the time difference between the 10\% and the 90\% of the 
leading edge, and the time difference between the 90\% and 30\% of the trailing edge, respectively.

\begin{table}[htb]
\centering
\caption{Technical details for the \ZNMO\ bolometers (cylinder and parallelepiped) and for the LD. 
The cylindrical \ZNMO\ was measured twice, so we reported the parameters also for the background run (Cylinder$^*$).
R$_{work}$ is the working resistance of the thermistors. Signal represents  the absolute voltage drop across the termistor for a unitary 
energy deposition.}
\label{Table:parameters_measurement}
\begin{tabular}{lccccc}
\hline
Crystal       &R$_{work}$    &Signal             &FWHM$_{base}$      &$\tau_{R}$    &$\tau_{D}$\\
              &[M$\Omega]$   &[$\mu$V/MeV]       &[keV]              &[ms]          &[ms]   \\
\hline
\hline
Cylinder      & 3.7          &140   			    & 0.6             & 17 			  & 50 \\
\hline
Parallel.     & 4.7          & 320     			    & 1.2             & 8  			  & 33\\
\hline				
LD            & 8.8          & 1700    				& 0.16            & 4 		      & 11 \\
\hline
Cylinder$^*$  & 2.5          & 200     			    & 0.7             & 17 		 	  & 48 \\
\hline
\end{tabular}
\end{table}

After the application of the optimum filter, signal amplitudes are  corrected for temperature and gain instabilities of the set-up thanks to
a monochromatic power injection in the Si heater taking place every few minutes.
The  \ZNMO\  is calibrated using the most intense $\gamma$-peaks from the \THO\ source, while the LD is calibrated using the 
$^{55}$Fe X-ray doublet.

The FWHM energy resolution obtained on the cylindrical (parallelepiped) crystal ranges from 2.5 $\pm$ 0.1 (2.4 $\pm$ 0.1) keV at 238 keV to 
3.8 $\pm$ 0.9 keV (7.6 $\pm$ 1.3) at 2615 keV. The energy resolution on the 5407 keV $\alpha$ + recoil line (due to a weak internal contamination  
of \PO\ ) can be evaluated only on the  long background run for the cylindrical crystal and gives 5.3~$\pm$~1.1~keV. 
The FWHM energy resolution on the LD, evaluated on the $^{55}$Fe X-ray doublet, is 321 $\pm$ 9 eV.
Experimental resolutions are worse than theoretical ones  in agreement with the observed performance of macro-bolometers~\cite{Bellini:2010iw}.

\begin{figure}[t]
\centering
\includegraphics[width=0.48\textwidth,clip=true]{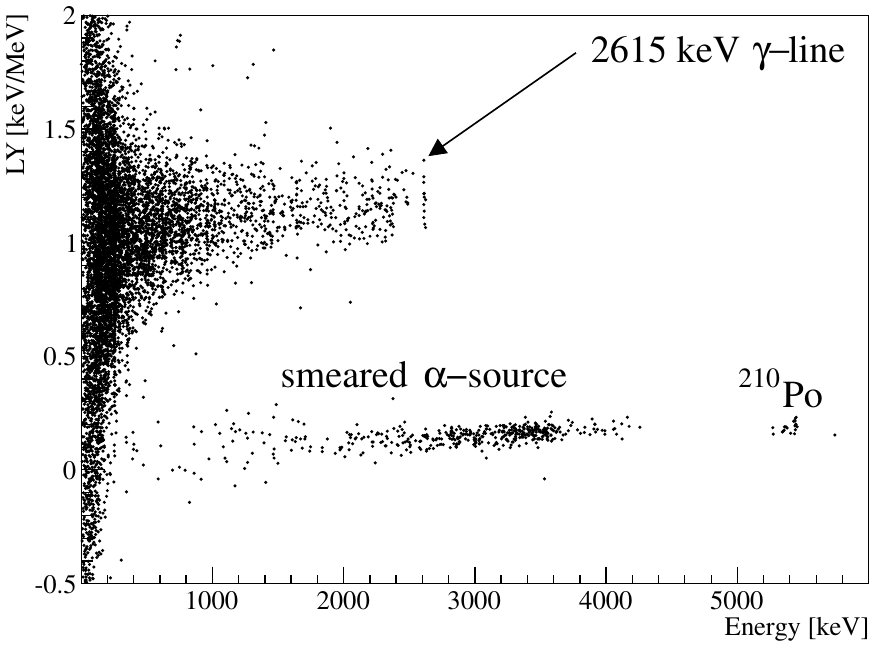}
\caption{The light-to-heat energy ratio as a function of the heat energy obtained with the cylindrical crystal in the first run, during a 62 h 
\THO\ calibration. The upper band 
(ascribed to $\beta/\gamma$ events)  and  lower band (populated by $\alpha$ decays) are clearly separated. The  2615 
keV \TL\ $\gamma$-line is well visible in the   $\beta/\gamma$ band as well as a the continuum rate induced by the 
degraded $\alpha$ source and the 5407 keV $^{210}$Po doublet in the $\alpha$  band. The discrimination power is reported in Sec.~\ref{sec:discrimination}.}
\label{fig:LY}   
\end{figure}

The light-to-heat energy  ratio as a function of the heat energy is shown for the calibration spectrum in Fig.~\ref{fig:LY}.
$\beta/\gamma$ and $\alpha$ decays give rise to very clear separate distributions. 
In the upper band, ascribed to  $\beta/\gamma$ events, the  2615 keV \TL\ $\gamma$-line is well visible. 
The lower band, populated by  $\alpha$ decays, shows  the continuum rate induced by the degraded $\alpha$ source as well as 
the  \PO\  doublet.

The Light Yield (LY), defined as the ratio between the measured light (in keV) and the nominal energy of the event (in MeV), 
was measured for the most intense $\gamma$-lines 
giving 1.10 $\pm$ 0.03 keV/MeV and 0.78 $\pm$ 0.02 keV/MeV for the cylinder and for the parallelepiped, respectively.
These values are constant from 0.2 to 2.6 MeV and are not corrected for the light collection efficiency.
The LY of the cylindrical crystal is well in agreement with the one reported in~\cite{Gironi:2010hs,Giuliani-2011},
while the parallelepiped shows a lower LY. 

The Quenching Factor (QF), defined as the ratio of the  LY$_{\alpha }$/LY$_{\beta/\gamma }$  
for the same energy release, was evaluated on the 5407 keV $\alpha$-line and results 0.18 for both  crystals.

\section{$\alpha$ vs $\beta/\gamma$  discrimination}
\label{sec:discrimination}
\begin{figure}[tbp]
\centering
\includegraphics[width=0.48\textwidth,clip=true]{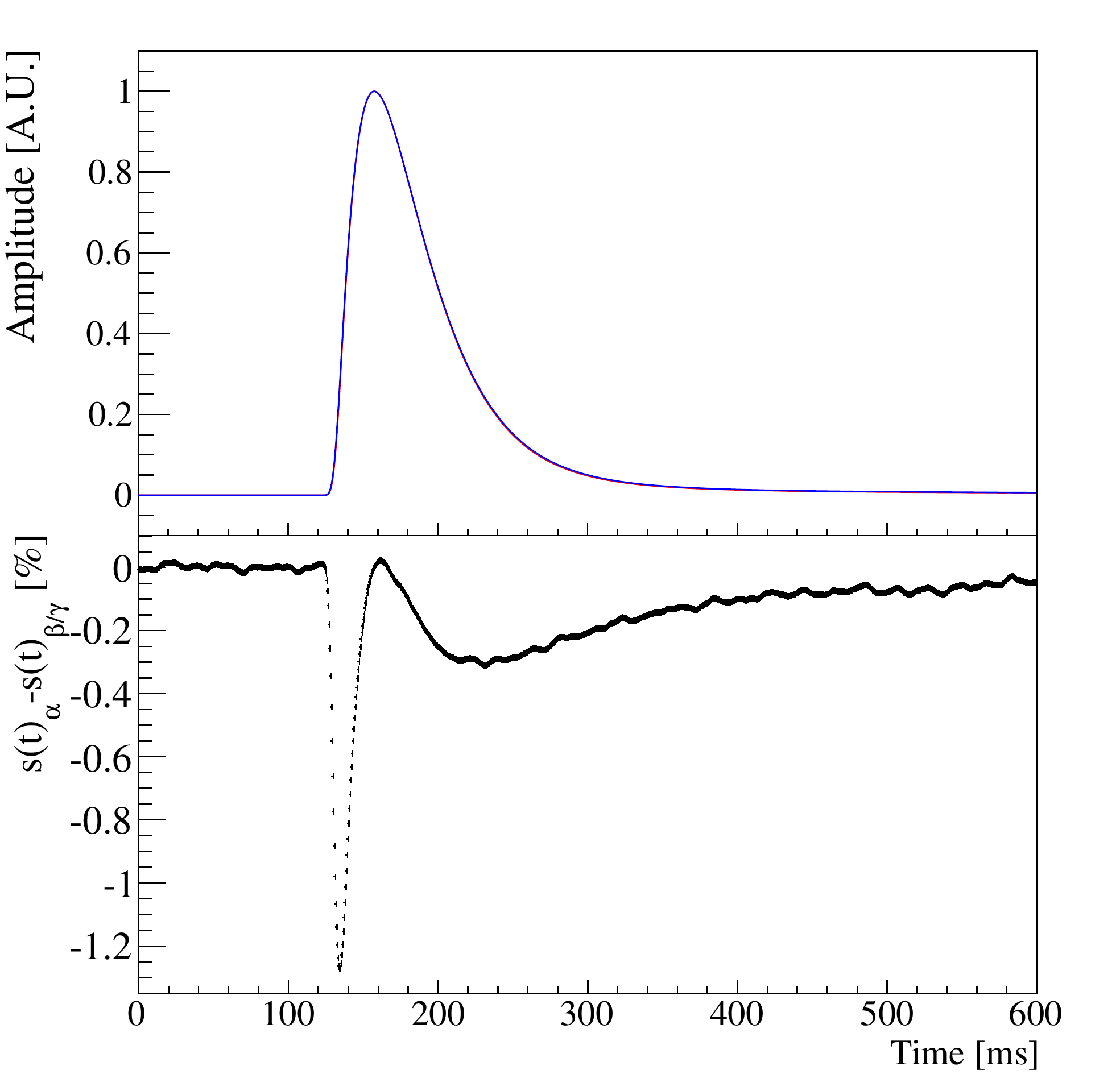}
\caption{Comparison of the $\beta/\gamma$ and $\alpha$ pulse shape. Top figure: thermal pulse obtained by averaging 
events in the 2615 keV $\gamma$ peak (blue line) and events belonging to the $\alpha$ band with an equivalent 
energy (red line): their difference is not appreciable.  Bottom figure: percentage difference $s(t)_{\alpha}$-$s(t)_{\beta /\gamma}$. 
Differences at a level of a few per mille are visible both in the rise and decay of the thermal pulse.}
\label{fig:average}      
\end{figure}

As reported in~\cite{Gironi:2010hs, Arnaboldi:2010gj}, Molybdate  crystals can provide $\alpha$~vs~$\beta/\gamma$ discrimination by making use 
of the thermal information only.
In Fig.~\ref{fig:average} the ideal signal shape $s(t)$ for the two event classes  is shown together with the percentage 
difference $s(t)_{\alpha}$-$s(t)_{\beta /\gamma}$. 
Pulse shapes are obtained by averaging pulses  (obtained in the same calibration measurement of Fig.~\ref{fig:LY}) in the energy range  2610-2620 keV  
and aligned at the maximum. 
Differences at a level of a few per mille are visible both in the rise and decay  of the thermal 
pulse~\footnote{We will refer to pulses  from the  cylindrical bolometer throughout the rest of  the text. 
However, the parallelepiped bolometer showed consistent results.}.

To fully exploit this feature several shape variables have been studied: $\tau_{R}$, $\tau_{D}$, TVL, TVR, $\rm{\chi^2_{OT}}$.
The Test Value Left (TVL) and Right (TVR) are defined as:

\begin{align}
{\rm TVL} &= \frac{1}{A\cdot w_l }\sqrt{\sum_{i=i_{M} - w_l}^{i_{M}}    \left(y_i -  A\, s_i\right)^2}; \\
{\rm TVR} &= \frac{1}{A\cdot w_r }\sqrt{\sum_{i=i_{M} }^{i_{M} + w_r} \left(y_i -  A\, s_i\right)^2}.
\end{align}

All quantities in the above equations refer to the optimum filtered pulses:   y$_i$ is  the pulse,  A and  i$_M$ its  amplitude and 
maximum position, s$_i$ the  ideal signal pulse scaled to unitary amplitude and aligned to y$_i$,  w$_{l(r)}$  the left(right) 
width at half maximum of s$_i$. The maximum of the ideal  pulse is aligned with the maximum of the filtered pulse, fractional 
delays are handled with a linear interpolation of the data samples.

Finally, $\rm{\chi^2_{OT}}$ was computed as the sum of the squared fit residuals normalized to the noise RMS, using  s$_i$ as 
a fit function of the central part of the pulse. This algorithm is described in detail in~\cite{DiDomizio:2010ph}.

Bolometers are nonlinear detectors~\cite{Vignati:2010yf}, i.e. the shape of the signal slightly depends on the amount of released energy.   
This  implies that also the shape parameters (evaluated with respect to an energy independent pulse, s$_i$) will show such a dependence.
To remove such a  dependency  variables were linearized in the 2300-3200 keV energy range.

\begin{figure}[h]
\centering
\includegraphics[width=0.48\textwidth,clip=true]{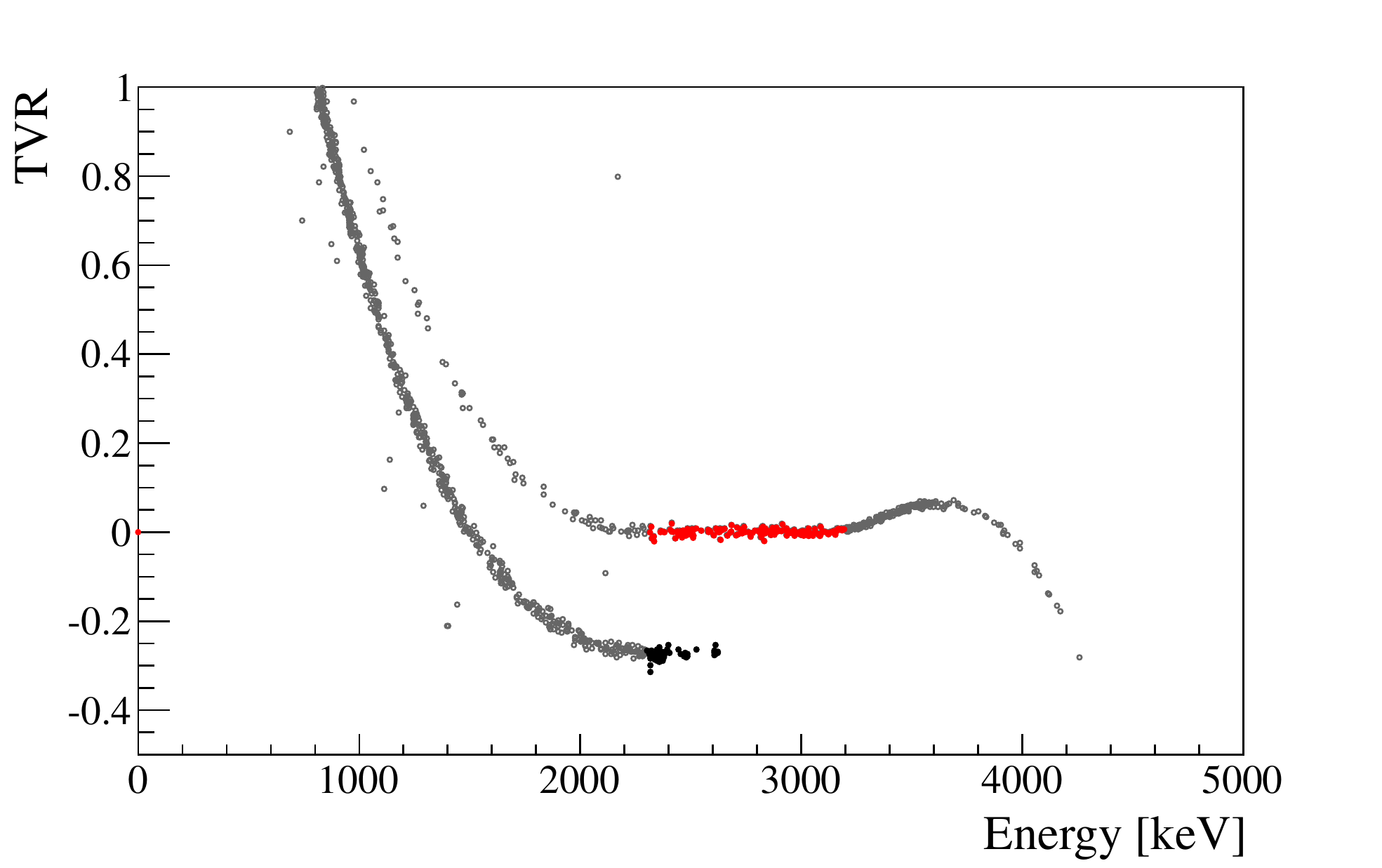}
\includegraphics[width=0.48\textwidth,clip=true]{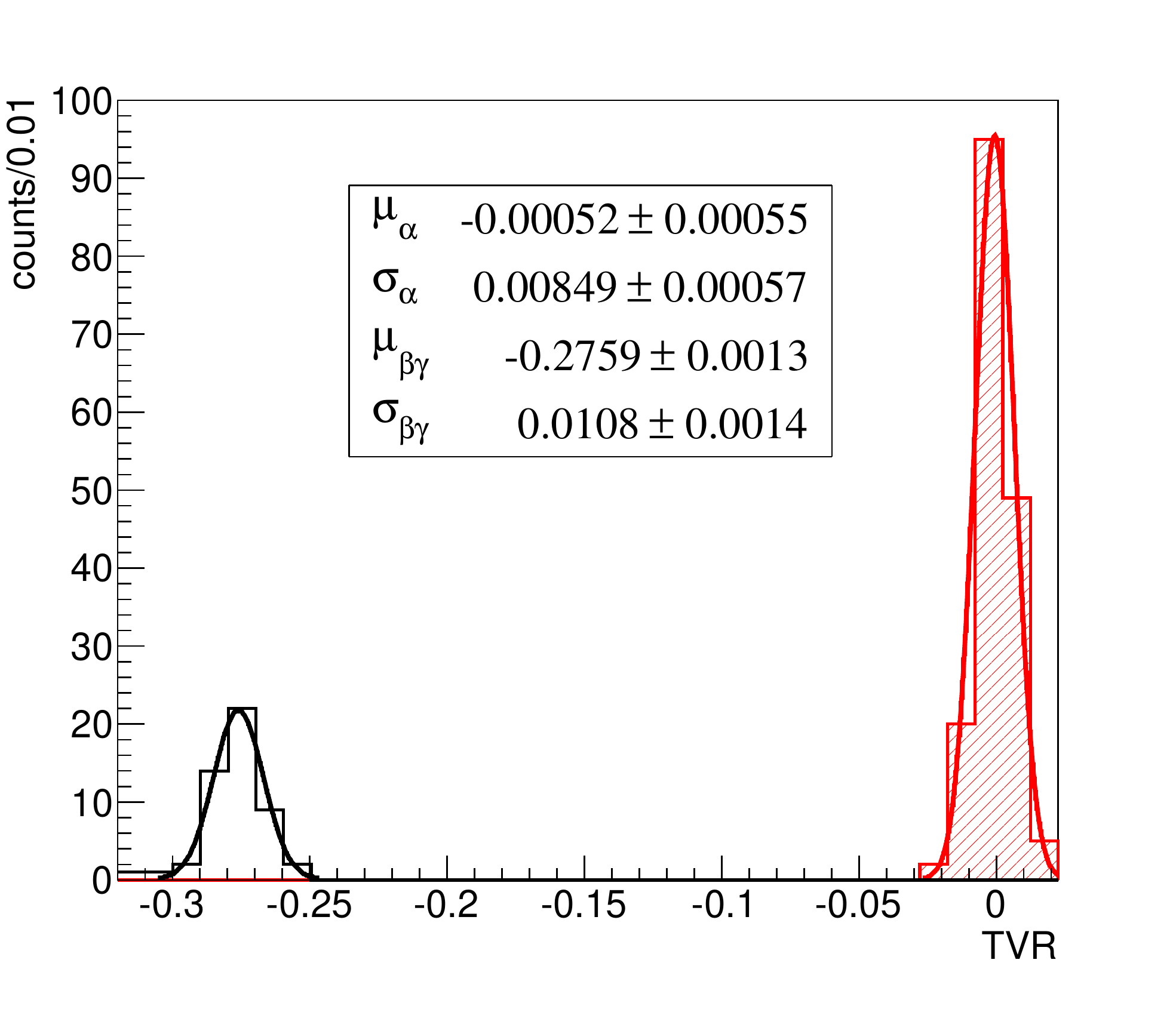}
\caption{Top Figure: TVR as a function of the energy. The upper band is populated by  $\alpha$ particles (events in 
the 2300-3200 keV energy range   are shown in red)  while  $\beta/\gamma$'s contribute to the lower band (events in the 
2300-3200 keV energy range  are shown in black). Bottom Figure:  TVR histogram for the $\alpha$ sample in red and for 
the $\beta/\gamma$ in black. The mean values and  the standard deviations, as estimated  from a Gaussian fit, are reported.  
Using Eq.~\ref{eq:DP} a discrimination power of $\approx$ 20 is obtained.}
\label{fig:Lin_TVR}      
\end{figure}
This procedure, moreover, allowed to enlarge the data sample on which the  pulse shape analysis was performed. 

Fig. \ref{fig:Lin_TVR} shows the   TVR variable as a function of the energy.  As for the case of  Fig.\ref{fig:LY},   
$\beta/\gamma$ and $\alpha$ events are distinctly separated.
Events in the linearized energy region and belonging to the upper ($\alpha$) band are shown in red while events, in the 
same energy interval, belonging to the lower ($\beta/\gamma$) band are shown in black.
The discrimination power (DP) between the two distributions is usually  quantified as  the difference between the average values of the two 
distributions  normalized to the square root of the quadratic sum of  their  widths: 
\begin{equation} 
\rm{DP} = \frac{\mu_{\beta/\gamma}-\mu_{\alpha}}{\sqrt{\sigma_{\beta/\gamma}^2+\sigma_{\alpha}^2}}.
\label{eq:DP}
\end{equation} 
The above definition  does not imply any probabilistic meaning, it was used for the sake of comparison with similar quantities reported  
in published papers~\cite{Gironi:2010hs, Arnaboldi:2010gj}.

The TVR distribution (see bottom panel in Fig.~\ref{fig:Lin_TVR}) exhibits a discrimination power of $\approx$ 20.
This is a very impressive result. For comparison, the Light Yield DP for the same events (Fig.~\ref{fig:LY}) is $\approx$ 8.
The separation achievable with the shape algorithm  alone  is therefore  much more powerful than the one obtained using the information of the 
light detector and  allows to reject $\alpha$ events to any  desirable level.

Results for all  the shape parameters are reported in Tab. \ref{tab:DP}.   Considering the outstanding  TVR performance compared to other 
variables, it was  not worth combining them into a multivariate analysis.
For completeness, TVR and scintillation  were combined~\cite{ Hocker:2007ht} into a Fisher linear discriminant analysis~\cite{Fisher,Fisher1}, resulting 
in a discrimination power of $\approx$ 24.
The two variables showed only a modest correlation: +23$\%$(-25$\%$) for the  $\beta/\gamma$($\alpha$) events.

\begin{table}[htb]
\begin{center}
\begin{tabular}{lcc}
\hline
\hline
Shape Variable      & DP\\
\hline
$\tau_R$      & 4.9 \\ 
$\tau_D$     & 4.6\\
TVL      & 4.2 \\
TVR     & 20.0\\
$\rm{\chi^2_{OT}}$ & 6.2\\
\hline
\end{tabular}
\end{center}
\caption{The discrimination power (DP) as defined in Eq.~\ref{eq:DP}  for all the shape variables.} 
\label{tab:DP}
\end{table}

\section{Internal contaminations}
\label{sec:Internal-contaminations} 
\begin{figure}[htb]
\centering
\includegraphics[width=0.48\textwidth,clip=true]{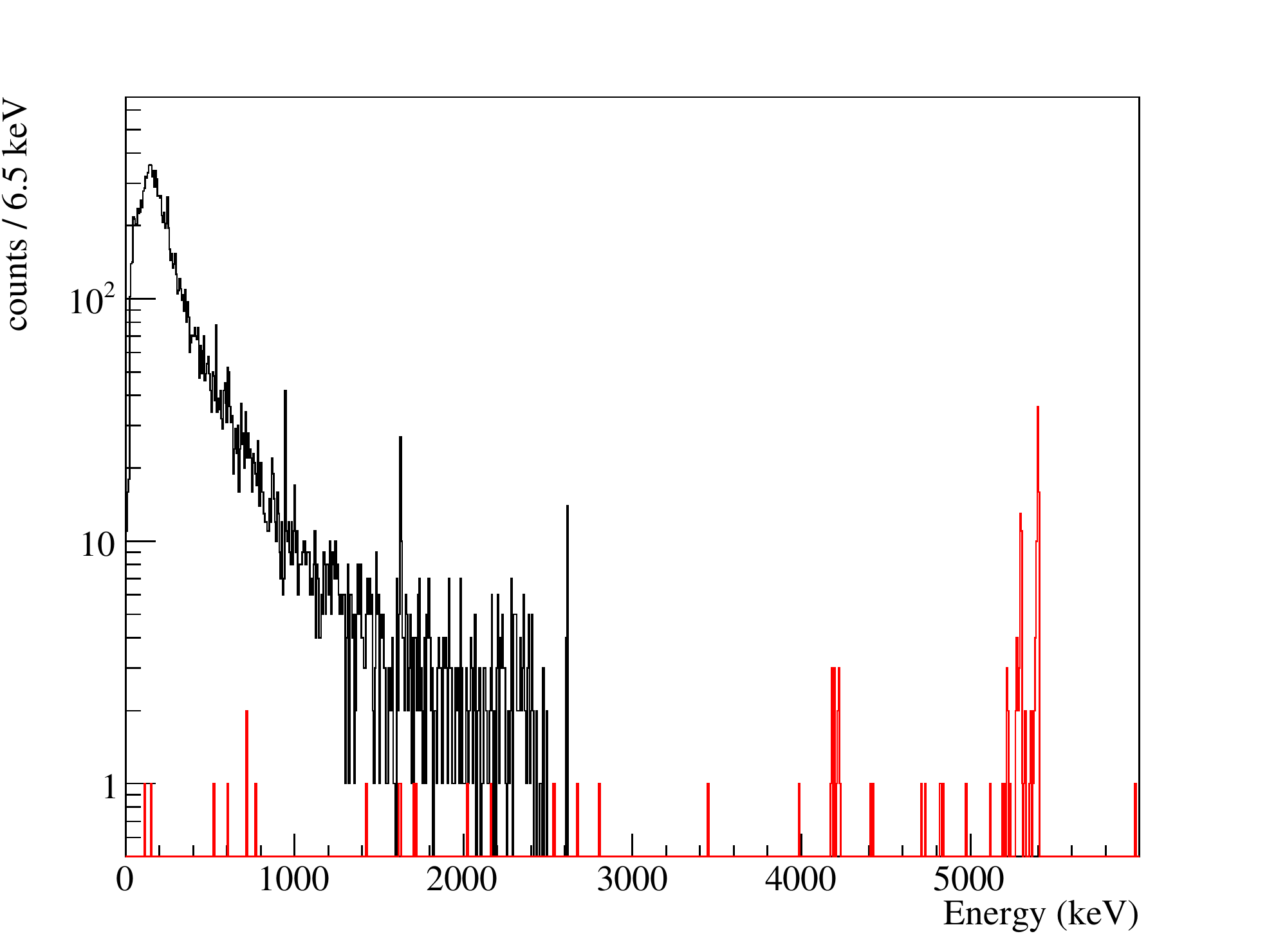}
\caption{Energy spectrum obtained in the second run with the cylindrical crystal corresponding to 407 hours of measurements. 
The spectrum includes also a \THO\ calibration. The $\alpha$ events, recognized through the TVR parameter, are in red.}
\label{fig:contamination}
\end{figure}

As pointed out in Sec.~\ref{sec:Experimental-details}, the internal contaminations were evaluated in a second, dedicated run. In this 
run only the cylindrical crystal was measured. 
The set-up of the detector was identical to the one  sketched in Fig.~\ref{fig:setup}, but this time the LD was not present. 
In order to avoid surface radioactivity, the Cu facing the crystal was covered by several layers of polyethylene, the same used in CUORE 
Crystal Validation Runs~\cite{CCVR-2011}.
The detector was run in similar temperature conditions of the previous run (see Tab.~\ref{Table:parameters_measurement}).
The obtained energy spectrum, consisting of a sum  of background and \THO\ calibration spectra (totalling 407 hours), is presented in  Fig.~\ref{fig:contamination}.

The $\alpha$ induced background  shows clearly three peaks. Two of them  are identified as  induced by  \PO\  (internal, 5407 keV, and on surface, 5304 keV) decay,
while the third one is due to an internal contamination of $^{238}\mathrm{U}$.
No other structure is visible on the spectrum, resulting in a flat continuum. In particular no events are recorded above 6 MeV, implying no decays induced 
by  the most dangerous radionuclides for DBD searches, namely the ones that undergo  $\beta$/$\gamma$ decay with Q-values larger than 
Q$_{\beta\beta}$ (see Sec.~\ref{sec:bulk-contamination}).
The evaluated internal radioactive contaminations are presented in Tab.~\ref{tab:contInterne}.

 \begin{table}
\begin{center}
\begin{tabular}{lcccc}
\hline\hline
Chain & nuclide   & activity\\
      &           & $\mu$Bq/kg \\ 
\hline
\THO\  & \THO\  & $<$ 32 \\
           & $^{228}$Th  & $<$ 32 \\
\hline
$^{238}$U  & $^{238}$U    & 350 $\pm$ 90 \\
           & $^{234}$U    & $<$ 110   \\
           & $^{230}$Th   & $<$ 68 \\
           & $^{226}$Ra   & $<$ 32 \\
           & $^{210}$Po   & 1660 $\pm$ 200\\
\hline\hline
\end{tabular}
\end{center}
\caption{Evaluated internal radioactive contaminations for the cylindrical crystal. Limits are at 68\% CL.} 
\label{tab:contInterne}
\end{table}

\section{Background simulation}
\label{sec:background-simulation} 
\subsection{General consideration}
The study of background issues for an experiment aiming at reaching high sensitivity on \DBD\ search can be divided into different 
types of sources depending on their position. In the following we will analyze  in detail the ``near sources'' (crystal and copper mounting 
structure contaminations). ``Far sources'' (radioactive contaminations in the experimental apparatus, mainly the refrigerator and its external 
shields) and ``environmental sources'' (contribution present at the experimental site: mainly muons, neutrons and $\gamma$ rays) are instead 
closely dependent on the experimental apparatus and environment in which the set-up is installed. A detailed description concerning these 
categories of sources in Cuoricino, the inspirer of this configuration  can be  found in~\cite{Bucci:2009fk}. 

Background induced by  far sources is produced  by $\gamma$-rays since $\beta$'s and $\alpha$'s can be easily shielded. 
The choice of an isotope with a Q-value exceeding 2615 keV (by far the most  intense -high energy- natural $\gamma$-line emitted 
by the decay of \TL\ , belonging to the \THO\  decay chain) allows to greatly reduce this background. 
Above this energy there are only extremely rare high energy $\gamma$'s from $^{214}$Bi: the total Branching Ratio (BR) in the energy window 
from 2615 up to 3270 keV is 0.15 \%. 

In this frame it is mandatory to take into account also coincidence events that can occur in one crystal. The most relevant case is 
represented by the decay of \TL\ .  In this decay, for example, there is 85 \% probability that the    2615 keV $\gamma$ is emitted in cascade with the
583 keV $\gamma$ . In fact, as we will evaluate in Sec.~\ref{copper-bulk-contamination}, this mechanism represents 
the largest source of background above 2615 keV. On the other hand it is clear that the probability of simultaneous interaction in the 
same crystal strongly depends on the relative distance from the detector. Moreover the lower energy $\gamma$ is easily shielded 
compared to the higher energy one, so that this type of background  can be completely neglected, in the case of  ``far sources'', 
with the installation of a proper shielding.
 
\begin{figure}
\begin{center}
\resizebox{0.25\textwidth}{!}
{\includegraphics{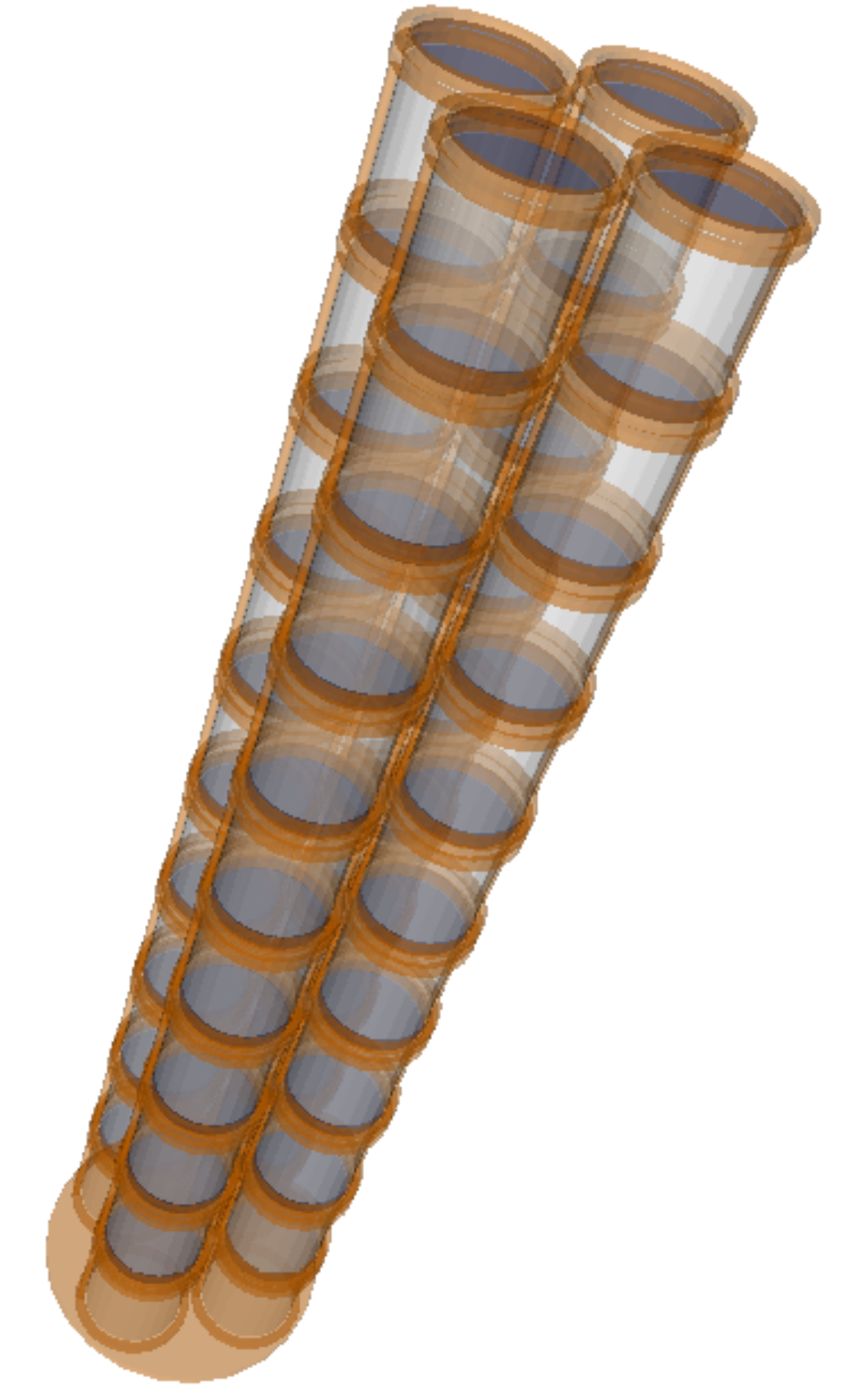}}
\caption{Geometry used for the simulation of the ZnMoO4 array: a tower-like detector of 40 cylindrical crystals 
(h = 60~mm, $\varnothing$ = 60~mm), where each  crystal is faced by one Ge LD (h = 0.5mm, $\varnothing$ = 60~mm).}
\label{fig:Torre}      
\end{center}
\end{figure}

Finally the last contribution that must be considered, for what concerns environmental sources, is the one coming from neutron and muon
interactions in the detector. For these we consider the results reported in~\cite{Bucci:2009fk,Andreotti:2009dk} by the \Cuore\ collaboration, for a 
modular \TEO\ bolometric detector operated underground, with a geometrical configuration similar to the one proposed  here (see Figure \ref{fig:Torre}). 
The change in the molecular compound does not heavily affect the background rates. Indeed, as in the case of \TEO\ crystals, none of the 
isotopes of \ZNMO\ have high neutron cross sections, the main parameter that could affect the background. External neutrons yield an integrated  
contribution  in the 3-4 MeV region $<$10$^{-5}$ \ckky\ ~\cite{Bucci:2009fk}, while external muons - operating the array in anticoincidence - yield a 
contribution of $<$10$^{-4}$ \ckky\ in the 2-4 MeV region~\cite{Andreotti:2009dk}.

\subsection{Near sources}

The region of interest (ROI) for the evaluation of the background is a few FWHM wide and centered at the \DBD\ Q-value. For \ZNMO\ crystals 
the ROI has therefore been chosen between 3014 and 3054 keV. All the background levels quoted in the following have been defined running 
dedicated \MC\ simulations based on \textsc{Geant4} code. A tower-like detector (see Fig.~\ref{fig:Torre}) was simulated, made out of 40 
detectors (\ZNMO\ crystal + Ge LD) arranged in a modular configuration of 10 floors with 4 crystals each. The \ZNMO\ crystals 
here considered are cylinders (h = 60~mm, $\varnothing$ = 60~mm) faced to Ge slabs of h = 0.5~mm, $\varnothing$ = 60~mm.

The total mass of the \ZNMO\ detectors is 29.3 kg, which corresponds to 1.20 kg of $^{100}$Mo in the case of non enriched crystals and to a 
mass of 11.22 kg in the case of enriched crystals at the 90\% level. The design of the detectors is done in such a way as to reduce the amount 
of inert material facing the detector and at the same time to overcome the problem of differential thermal contractions between the 
detector and  structure. The most ``massive'' materials employed for the detectors assembly are copper (skeleton structure) and 
PTFE (detector holders), and the mass ratio between these two components is about 100:2. 
The contribution of PTFE is neglected due to the reduced mass and relatively high radio-purity of this material~\cite{Aprile:2011ru}.   
Thus the only relevant near sources of background 
are radioactive contaminations of the copper  structure and of the scintillating crystals themselves. In both cases not 
only $\gamma$'s, but also $\beta$'s  contribute to the counting rate in the \DBD\ region. 

The contamination of the Ge wafers, if any, will not be a problem: I) high purity Ge slab 
will be used as LD's with negligible radioactive internal contaminations,  II) radioactive decay 
in the LD produce thermal pulses with a different shape than events produced by scintillation light. These events can 
therefore be easily recognized and rejected.

In the following sections, all the possible near sources will be considered separately.

\subsubsection{Crystal bulk contaminations}
\label{sec:bulk-contamination}
In the case of crystal bulk contaminations the main contribution to the background rate in the ROI arises from the $\beta$/$\gamma$ emissions of 
the natural radioactive chains. Indeed the kinetic energy of $\alpha$'s emitted by U and Th daughters are always far above the ROI. The $\beta$ emitters that 
give an important contribution to the background are \TL\ (\THO\ chain), $^{214}$Bi and $^{210}$Tl ($^{238}$U chain). 

\TL\ has a Q-value of 5 MeV and decays through a $\beta$ channel with a half-life of 3 min. It gives rise to a continuous spectrum 
which, in combination with $\gamma$'s emitted in the re-assessment of the daughter nucleus, can produce events falling in the ROI. However, the background induced by 
this isotope can be effectively reduced by means of delayed coincidences between the \TL\ signal and the $\alpha$ emitted by its 
precursor, $^{212}$Bi (Q-value=6.2 MeV). The coincidence pattern to be studied would be an event in the \DBD\ region preceded by 
an $\alpha$ decay of 6.2 MeV. Provided that the $\alpha$ counting rate is not too high, the dead time introduced by this method
is low or even negligible.

In the $^{238}$U decay chain the only contribution above 2.7 MeV comes from the $\beta$ decay of $^{214}$Bi and $^{210}$Tl. 
$^{214}$Bi decays with a BR of 99.98\%  (Q-value=3.27 MeV) to $^{214}$Po which  $\alpha$ decays   with a very short half-life (163 $\mu$s) and 
a Q-value of 7.8 MeV. Since the rise time in our bolometers is of the order of ms, the $\beta$ events add up to $\alpha$-events producing 
events  with energies higher than 7.8 MeV (the so called Bi-Po's). These events therefore do not produce background into the ROI.
In the residual 0.02\% of cases, $^{214}$Bi $\alpha$-decays to $^{210}$Tl that is a $\beta$ emitter with a Q-value of 5.5 MeV 
and an half-life of 1.3 min. The background induced by this $\beta$ emitter can be easily rejected with a cut on the delayed 
coincidences between the $^{210}$Tl signal and the $^{214}$Bi $\alpha$, as it happens for $^{212}$Bi.

For the sake of completeness we neglect, at this level,  internal contaminations of ``exotic'' long lived, high 
energy (Q$>$3 MeV, $\tau$$>$ 0.3 years) $\beta$ emitters, the most dangerous being 
$^{42}$Ar ($^{42}$K), $^{106}$Ru ($^{106}$Rh) and $^{126}$Sn ($^{126}$Sb).

\subsubsection{Copper bulk contamination}
\label{copper-bulk-contamination}
Copper is the  most abundant material in the experimental set-up near the detectors. The most important contribution to 
the counting rate in the ROI might be given by $^{214}$Bi high energy $\gamma$'s, and by \TL\ $\gamma$'s at 583 and 2615 keV 
that, being emitted in cascade, can sum up to 3.2 MeV.\newline A proper selection of high quality copper would be enough to guarantee 
a negligible contribution from these sources. 
Indeed, simulating a bulk contamination of 19 $\mu$Bq/kg in $^{228}$Th and 16 $\mu$Bq/kg in $^{226}$Ra (limits on NOSV copper ~\cite{Laubenstein}) 
in the detector copper structure, the overall contribution in the ROI is 2.40$\pm$0.04$\cdot$ 10$^{-4}$ \ckky\ .

Considering the same bulk contaminations for the copper of the 10 mK shielding (the inner thermal shielding, 2 mm thick) the total background
expected is 1.40$\pm$0.05 $\cdot$  10$^{-4}$ \ckky . 
The background due to the other, more distant, thermal shielding and the other far parts of the cryostat is  lower 
and  therefore, negligible.

\subsubsection{Surface contaminations}

An important contribution to the background of the experiment may arise from surface contamination, specially from degraded $\alpha$ coming 
out mostly from surfaces of inert materials facing the detector~\cite{ACryo,Pavan:2008zz}. 
Two procedures can be applied  for its reduction:  $\alpha$ rejection based on pulse shape analysis (or the scintillation signals) and 
the use of anticoincidences between the detectors of the array. This last possibility is, however, useful only for crystals surface 
contaminations. It does not allow in any way to recognize the $\alpha$ background due to surface contamination of inert material facing
the detectors. Taking into account the discrimination power obtained both with pulse shape analysis and scintillation signals 
described in Sec.~\ref{sec:discrimination}, this background source can be completely neglected.

\subsubsection{$\beta$/$\gamma$ pile-up}
\label{sec:pileup}
Finally, one has  to take into account that for slow detectors, like bolometers, even a feeble $\beta$/$\gamma$ emission with Q$<$3 MeV can 
produce an unwanted background due to the difficulty to recognize and reject pile-up on the rise time of the thermal pulse. Indeed 
independent signals can randomly add generating a background that extends even beyond the ROI.

This background depends mainly on 3 factors: the time within which is impossible to distinguish two separate events, the rate of 
$\beta$/$\gamma$ interactions and their energy distribution. With respect to the rate due to $\beta$/$\gamma$ events arising from near sources or internal 
contamination, it can be easily evaluated given the contaminations of Sec.~\ref{sec:bulk-contamination} and Sec.~\ref{copper-bulk-contamination}.
This rate is of the order of 0.1 mHz. Even considering the far sources (that are negligible in the ROI but contribute in the 0--2615 keV region) 
the total rate is well below 1 mHz.
 
In fact it turns out that the main contribution arises from  the 2$\nu$DBD of $^{100}$Mo, due to its relatively ``fast'' decay time. 
The expected rate in a 90\% enriched crystal (h = 60~mm, $\varnothing$ = 60~mm)  is 5.45 mHz. 
Unfortunately,  the energy spectra of the 2$\nu$DBD exhibit a maximum at rather high energy ($\simeq$ 1.25 MeV) with a ``broad'' 
distribution, so that the pile-up energy spectrum distribution falls very close to the 0$\nu$-ROI.

The pile-up is recognized through the pulse shape analysis of the optimally filtered pulses. The time window in which we are not able to distinguish two pulses
depends mostly on two parameters: the noise of the baseline and the relative pulse height of the two pulses. 
Our simulation (involving a dedicated pulse shape and noise generator) 
shows that the time in which two pulses cannot be recognized is  definitely smaller  than $\tau_{R}$/10. Here we conservatively assume 
as 5 ms the time in which we are not able to distinguish two different pulses.
Giving this as input, we performed a \MC\ simulation of the 2$\nu$DBD of $^{100}$Mo. The irreducible 2$\nu$DBD pile-up evaluated in the ROI is 
$1.96\pm 0.36 \cdot 10^{-3}$\ckky\ .

\subsection{Total background evaluation}
\label{sec:BackEval}

In this section we report the background level that could be reached taking into account the above mentioned sources. 
When quoting background rates, we assume to operate the detectors in anticoincidence, thus recording only single hit events.
Background contributions due to sources described in Sec.~\ref{sec:bulk-contamination}, Sec.~\ref{copper-bulk-contamination}, and Sec.~\ref{sec:pileup} 
are reported in the last column of Tab.~\ref{tab:fondo}. 
\begin{table}
\begin{center}
\begin{tabular}{lcccc}
\hline\hline
source      & position         & background\\
            &                  & [\ckky\ ]\\
\hline

U chain     & crystals bulk    	 	& 	 $<$ 1.16$\pm$0.1  $\cdot$ 10$^{-5}$ \\
Th chain    & crystals bulk    		& 	 $<$ 2.18$\pm$0.04 $\cdot$ 10$^{-4}$\\
2$\nu$DBD pile-up  & crystals bulk  & 	 		1.96$\pm$0.36 $\cdot$ 10$^{-3}$\\
U+Th chains & Cu frame bulk    		& 	 $<$ 2.40$\pm$0.04 $\cdot$ 10$^{-4}$\\
U+Th chains & Cu  shield bulk  & 	 $<$ 1.40$\pm$0.05  $\cdot$ 10$^{-4}$\\
\hline\hline
\end{tabular}
\end{center}
\caption{Background projection for a \ZNMO\ tower-like experiment, as proposed in Fig.~\ref{fig:Torre}. Contaminations in Cu refer 
to the mounting structure. The background reported for internal contaminations in U and Th chains takes into account the 
background suppression obtained with delayed coincidences at 5~$\tau$ (i.e. $\approx$9 min for the U chain and $\approx$22 min for the 
Th chain). The dead time therefore introduced is 0.01\% for U chain and 2.0\% for Th chain. The background due to the 2$\nu$DBD 
of $^{100}$Mo  reported here corresponds to 90\%  enriched crystals. Only statistical errors are quoted.} 
\label{tab:fondo}
\end{table}

The background reported for internal contaminations in U and Th chains takes into account the  suppression obtained 
with delayed coincidences at 5 $\tau$. The dead time therefore introduced is 0.01\% for U chain and 2.0\% for Th chain. 
The inefficiency introduced by those  events where the $\alpha$ decay occurs near the crystal surface and 
the $\alpha$ particle escapes the crystal is of about 0.1\%. This evaluation is based on \MC\ simulations. 
In this case we assume to loose any possibility of delayed coincidence, and we compute the background as the sum of the 
fraction of the untagged $\beta$ events  due to the $\alpha$ escape and the fraction of the untagged $\beta$ events  because 
they are out of the maximum allowed delayed coincidence window. 

\begin{figure}
\begin{center}
\resizebox{0.48\textwidth}{!}
{\includegraphics{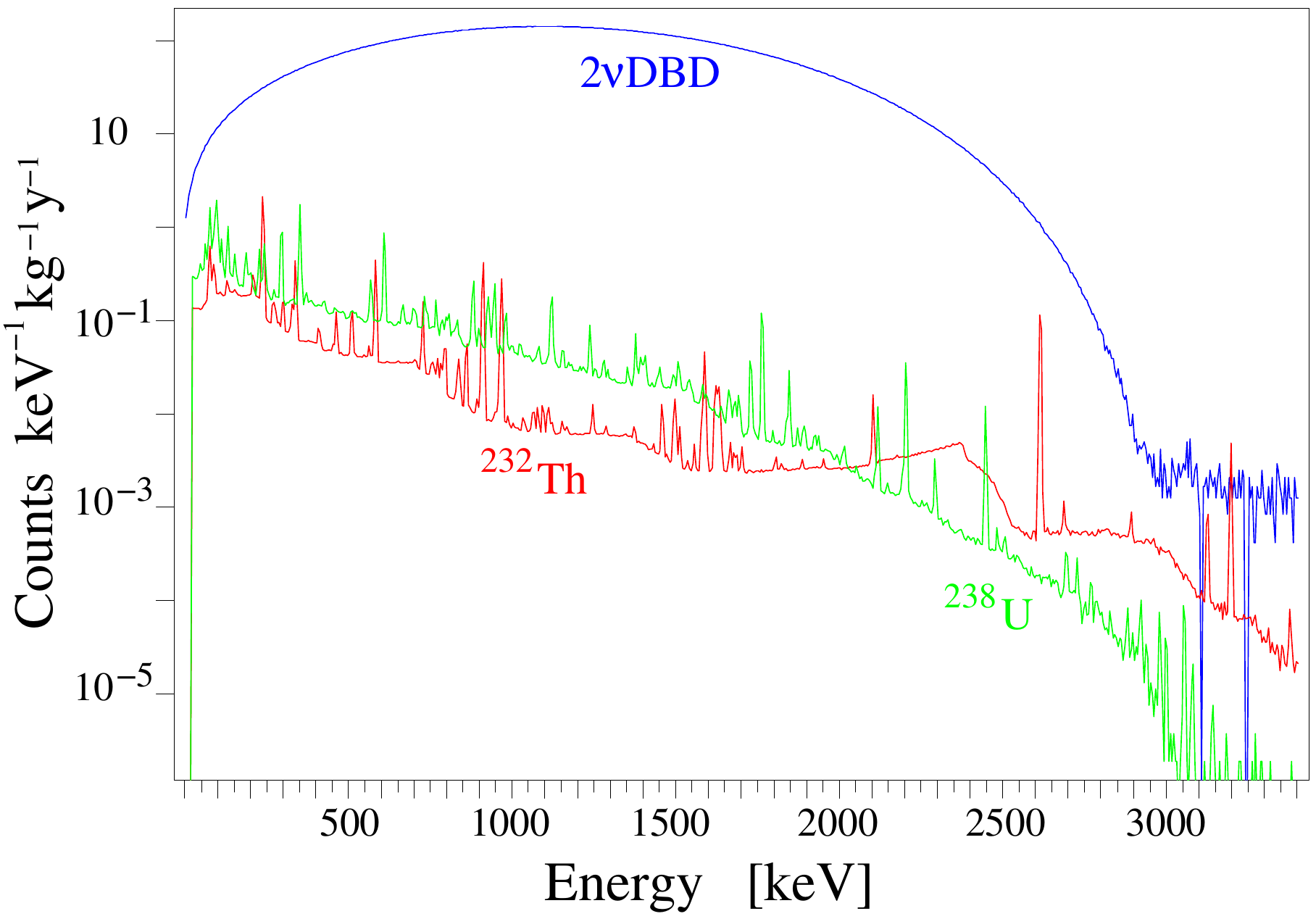}}
\caption{Simulated background spectra induced by \THO\ (red) and $^{238}$U (green) contamination in the Cu of the frames. 
The one induced by the 10~mK Cu shield is similar in shape, but a factor $\simeq$ 2 smaller. 
The background induced by the unrecognized 2$\nu$DBD pile-up (blue) clearly dominates.}
\label{fig:simulated_background}      
\end{center}
\end{figure}

Contribution due to the other ``near sources'' (i.e. the Cu frames and the Cu 10 mK shield bulk contaminations) are evaluated 
considering both the $^{214}$Bi high energy $\gamma$-lines and the \TL\   $\gamma$'s cascade. 
The induced background spectra due to \THO\ and $^{238}$U in the Cu frames is reported in Fig.~\ref{fig:simulated_background}.

As can be seen from  Tab.~\ref{tab:fondo} and from Fig.~\ref{fig:simulated_background}, the main contribution to the background is definitely due 
to unrecognized 2$\nu$DBD pile-up of $^{100}$Mo.  The only way to reduce this background will be to reduce the size of a single 
crystal but, as we shall see shortly, this would be useless for an experiment with a mass of 11.22 kg of $^{100}$Mo such as the one here investigated.

\subsection{Experimental sensitivity}
\label{sec:ExpSens}

The maximum sensitivity reachable by a \DBD\ experiment corresponds  to the ``zero background'' condition. This occurs 
when ($B \cdot M \cdot T \cdot \Delta) \simeq$ 1~\cite{Cremonesi2011}, where $B$ is the background level per unit 
mass, energy, and time, $M$ is the detector mass, $T$ is the measuring time and $\Delta$ is the FWHM energy resolution. 
For a ``standard'' live-time for a \DBD\ experiment (i.e. 5 years), an energy resolution of 5 keV and a detector 
mass of $\sim$30 kg the  ``zero background'' condition will be reached with a background of $\approx$1.5$\cdot$10$^{-3}$ \ckky\ .

Considering the different sources of background discussed in the previous sections and summarized 
in Tab.~\ref{tab:fondo}, with the detector array described above, for an enrichment of 90\%, in 5 years of data taking we 
are able to fulfill the ``zero background'' condition. 

For such an experiment, for a neutrino mass of   $<$~$\!m_{\beta\beta}\!>=100~meV$ 
we will expect between 1.7 and 13.9 counts of \DBD\ depending on the Nuclear Matrix 
Element~\cite{PhysRevC.77.045503, Civitarese:2009zz, Menendez2009139, PhysRevC.79.044301}.

\section{Conclusions}
We successfully tested two \ZNMO\ crystals as bolometers. The separation achievable on the shape of the thermal signal  alone  is  much more powerful 
than the one obtained using the information of the light detector and  allows to reject $\alpha$ events to any  desirable level.
Furthermore we would like to point out a very important consideration in favour of the pulse shape discrimination.
The light collection generally  depends on the size of the scintillating crystal, due to self absorption mechanisms. 
Since the LY of the  \ZNMO\ is rather small, this could imply a decrease in the achievable discrimination power obtainable 
through the light detection, moving from a few tenth  grams crystals to a few hundreds grams crystals. 
On the contrary, this mechanism will enhance the pulse shape discrimination since the self absorbed light signal will convert into heat, summing up with 
the non-radiative  de-excitations that makes possible the $\alpha$~vs~$\beta/\gamma$ discrimination.

Moreover, even without any kind of material selection, the internal contaminations in  $^{228}$Th and $^{226}$Ra are already at an 
extremely low level, never obtained  in any crystal compound based on Molybdenum.
The projection of these results to a small-size experiment foresees a possible background level close to zero.

\section {Acknowledgements}
This work was made in the frame of the LUCIFER experiment, funded by the European Research Council under the European Union's Seventh Framework 
Programme (FP7/2007-2013)/ERC grant agreement n. 247115. 
Thanks are due to E. Tatananni, A. Rotilio, A. Corsi, B. Romualdi and F. De Amicis  for continuous and constructive help in the  overall 
set-up construction.

\end{document}